\documentclass[%
 reprint,
nofootinbib,
nobibnotes,
bibnotes,graphicx,graphics,
 amsmath,amssymb,
 aps,stmaryrd,
]{revtex4-1}
\usepackage{natbib} 
\usepackage{soul}
\newcommand{\comment}[1]{}
\usepackage{tikz}
  \newlength\squareheight
  \newcommand\squareslash{\tikz{\draw (0,0) rectangle (\squareheight,\squareheight);\draw(0,0) -- (\squareheight,\squareheight)}}
  
  \newcommand{\vin}{\rotatebox[origin=c]{-45}{$\squareslash$}}
  
  \newcommand{\cin}{\rotatebox[origin=c]{+45}{$\squareslash$}}

\usepackage{wasysym} 
\usepackage{graphicx}
\usepackage{dcolumn}
\usepackage{bm}

\begin{document}

\preprint{Draft}

\title{Random graphs with arbitrary clustering and their applications}

\author{Peter Mann}
\email{pm78@st-andrews.ac.uk}
\author{V. Anne Smith}%
\author{John B.O. Mitchell}
\author{Simon Dobson}
\affiliation{School of Computer Science, University of St Andrews, St Andrews, Fife KY16 9SX, United Kingdom }
\affiliation{School of Chemistry, University of St Andrews, St Andrews, Fife KY16 9ST, United Kingdom }
\affiliation{School of Biology, University of St Andrews, St Andrews, Fife KY16 9TH, United Kingdom }

\date{\today}

\begin{abstract}
The structure of many real networks is not locally tree-like and hence, network analysis fails to characterise their bond percolation properties. In a recent paper [P. Mann, V. A. Smith, J. B. O. Mitchell, and S. Dobson,“Percolation in random graphs with higher-order clustering,”arXiv e-prints, p. arXiv:2006.06744, June 2020.], we developed analytical solutions to the percolation properties of random networks with homogeneous clustering (clusters whose nodes are degree-equivalent). In this paper, we extend this model to investigate networks that contain clusters whose nodes are not degree-equivalent, including multilayer networks. Through numerical examples we show how this method can be used to investigate the properties of random complex networks with arbitrary clustering, extending the applicability of the configuration model and generating function formulation. 
\end{abstract}

\pacs{Valid PACS appear here}
\maketitle


\section{Introduction}
\label{sec:introduction}
The properties of networks have proven of great interest to the statistical physics community \cite{newman_strogatz_watts_2001,newman_2019, cohen_havlin_2010,dorogovtsev_mendes_2013}, finding applications across biological, social and online networks to name a few. Networks have the ability to model the governing topological dynamics through simple models of nodes and edges. 
The central, information-rich object of a network is its degree distribution, $p(k)$, which is the probability distribution of picking a node from the network that has precisely $k$ edges. The degree distribution has a significant influence on the structural properties of the network, such as the robustness, path lengths, clustering and the dynamics of spreading processes such as percolation \cite{newman_2019}. 

Bond percolation is a binary-state process that considers the networks edges as either occupied or empty with a given probability, $\phi$. At some critical probability, $\phi_c$, a giant connected component (GCC) forms among the occupied edges through a 2nd-order phase transition. The position of the critical bond occupation probability and the macroscopic properties of the network are determined by local topological properties of the nodes and their degree distribution. The percolation properties of a network, including the onset of the formation of the GCC and its size as a function of $\phi$, are important quantities that are shaped by the topology and connectivity of the graph. 

A long-standing problem in network science is the rationalisation of the bond percolation process over a network with dense clustering \cite{PhysRevE.97.052306,PhysRevE.81.066114,PhysRevLett.97.088701,PhysRevE.68.026121,gleeson_2009,benson_gleich_leskovec_2016,fronczak_holyst_jedynak_sienkiewicz_2002,hebert-dufresne_noel_marceau_allard_dube_2010,PhysRevE.97.052306,allard_hebert-dufresne_young_dube_2015,karrer_newman_2010,PhysRevE.68.026121,PhysRevE.81.066114,PhysRevE.93.030302}. Clustering is defined as the failure of the graph to be tree-like; there exist edges connecting neighbours of nodes together in a cycle. Since real networks will almost certainly contain cycles at some order, model networks often fall short in describing the properties of real networks. Newman and Miller independently developed a generating function formulation that enabled the study of networks containing tree-like and closed triples \cite{PhysRevLett.103.058701,PhysRevE.80.020901}. Additional research to extend these models was also conducted in \cite{karrer_newman_2010,allard_hebert-dufresne_young_dube_2015}. These generating function models are based on joint degree distributions that partition the edges of a node into either tree-like or triangle motifs. There have also been recent advances in this endeavour using the related method of message passing \cite{newman_2019, cantwell_newman_2019}.

In a recent paper \cite{2020arXiv200606744M}, we studied the percolation problem for random networks that exhibit homogeneous clustering, (clusters whose nodes are degree-equivalent to one another), to any-order by extending the tree-triangle formulation of Newman and Miller. Examples of homogeneous clusters include cliques or skeleton cycles with no interior edges between cycle-nodes. Additionally, a class of cycles can be formed by the successive weakening of cliques through the removal of an interior edge from each node-pair, such that all sites in the resulting structure have the same degree. Our main result was an analytical formulation that expresses the fraction of the network occupied by the GCC during the bond percolation process in clustered networks. While shown to be inexact, the magnitude of the error of the approximation can not be seen until the scale is magnified by at least three orders of magnitude. We also found the generalisation of the Molloy-Reed criterion for these complex systems \cite{molloy_reed_1995}.

This framework uses the configuration model to generate the random networks \cite{newman_strogatz_watts_2001,fosdick_larremore_nishimura_ugander_2018}. In the tree-triangle configuration model, nodes are assigned tree-like and triangle degree stubs. The node stubs are then connected together at random. In the large, sparse limit, these networks can be though of as graphs drawn uniformly at random from the set of all graphs whose nodes have identical joint degree sequence. For a large number of nodes, the probability of forming a triangle, or other higher-order cycles by accident during construction becomes vanishingly small. Additionally, the probability of multiple edges between the same two nodes is vanishingly small in this limit. As such, the structural properties of a configuration model network can be described accurately with the generating function formulation \cite{newman_strogatz_watts_2001, karrer_newman_2010, PhysRevLett.103.058701, PhysRevE.80.020901}. 

In this paper, we generalise this approach to study clustered networks with inhomogeneous cycles. In these networks, the sites within a cluster motif are not necessarily equivalent.  Importantly, this generalisation allows us to study multilayer networks with inter- and intra-layer clustering; since, nodes in different layers are non-equivalent. We present our model by considering interesting examples of random networks containing inhomogeneous clustering arising through different means. Importantly, we then describe how the percolation properties can be extracted algorithmically from a network subgraph and present sample code in the supporting information. In this paper, we generate our graphs according to the configuration model described above; however, we can assign nodes stub degrees from a vector of topologies $\bm \tau=\{\bot,\Delta,\square,\dots\}$, where $\bot$ represents single edges, $\Delta$ represents triangles and so on.

\section{inhomogeneous clusters}
\label{sec:inhomogeneous}

The generating function formulation \cite{Newman2002SpreadOE,newman_strogatz_watts_2001} describes the macroscopic properties of the collective system in terms of the local environment of a particular focal node of particular degree, averaged over the distribution of all degrees. In the generalised model \cite{PhysRevLett.103.058701,PhysRevE.80.020901,karrer_newman_2010,2020arXiv200606744M} the local environment surrounding a node is partitioned into tree-like and triangular degrees and the probability, $g_\tau(\phi)$, of remaining unattached to the GCC as a function of bond occupancy probability, $\phi$, is formulated for each cycle topology $\tau\in \{\bot, \Delta\}$, where $\bot$ symbolises tree-like and $\Delta$ symbolises triangle motifs. Each $g_\tau$ is constructed in terms of $u_\tau$, the probability that a node with a $\tau$-cycle is not part of the GCC. Since all nodes in cliques and simple or \textit{weak} cycles (and classes of successively weakened clique whereby all nodes have the same degree) are equivalent through symmetry, their $u_\tau$ probabilities are also identical.  

The consideration of higher-order cycles beyond $\bot$ and $\Delta$ was first considered by Karrer \textit{et al} \cite{karrer_newman_2010} and Allard \textit{et al} \cite{allard_hebert-dufresne_young_dube_2015}. In our recent paper, we introduced an analytical expression for $g_\tau$ to consider additional, homogeneous, cycles other than $\bot$ and $\Delta$ and introduced a vector, $\bm \tau=\{\bot,\Delta,\dots,\omega\}$, of topological cycles the local neighbourhood of a node can be decomposed into. To this vector we ascribed a joint degree distribution $p(k_\bot,k_\triangle,\dots k_\omega)$ that describes the probability of choosing a node at random from the network which has precisely $k_\bot$ tree-like edges, $k_\triangle$ edges involved in triangles and so on. This information-rich object is paramount to the analytical formulation and can be used to elucidate the percolation properties of the networks they describe.

To solve for $g_\tau$ we must first find $u_\tau$. To do this we form self-consistent equations for $u_\tau$ using the generating function of the excess $\tau$-degree as 
$u_\tau=G_{1,\tau}(g_{\bot},g_\Delta,\dots,g_\omega)$ for each $\tau\in \bm \tau$ where $G_{1,\tau}(\bm z) $ is defined \cite{2020arXiv200606744M} as
\begin{equation}
    G_{1,\tau}(\bm z) = \frac 1{\langle k_\tau\rangle}  \frac{\partial G_0}{\partial z_\tau}\label{eq:Jacobian}
\end{equation}
which describes the distribution of cycles found by following an edge in a $\tau$-cycle to a node. The probability that a fraction, $S$, of the network is occupied by the GCC is then $S=1-G_0(g_{\bot},g_\Delta,\dots,g_\omega)$, where $ G_0(\bm z)$ is defined by
\begin{equation}
   \sum^\infty_{k_\bot=0}\sum^\infty_{k_\triangle=0}\cdots \sum^\infty_{k_\omega =0}p(k_\bot,k_\triangle,\dots,k_\omega)z_\bot^{k_\bot}z_\triangle^{k_\triangle}\cdots z_\omega^{k_\omega}
\end{equation}
or, using a condensed notation
\begin{align}
     G_0(\bm{z})=& \sum_{k_\tau\in \bm{\tau}}^\infty p(\bm{k_\tau})\bm{z}^{k_{\bm{\tau}}}\label{eq:G0}
\end{align}
for $\bm \tau=\{\bot,\triangle,\dots,\omega\}$. It generates the probability of choosing a node with a given joint degree sequence over all permissible combinations in the network. 

The problem formulation therefore relies on the definition of the vector of topologies into which the network is decomposed. Given that the nodes in a homogeneous cluster are equivalent, it is sufficient to represent these cycles only once in $\bm \tau$. To study inhomogeneous cycles, we must consider each unique site in a particular cluster as a different cycle; since, each $g_\tau$ equation will be distinct for a given site in the cluster. 
\begin{figure}[ht!]
\begin{center}
\includegraphics[width=0.5\textwidth]{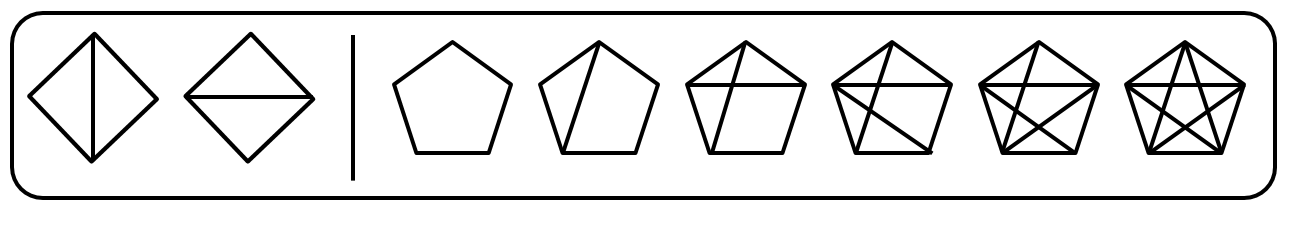}
\caption[diamond]{ 
(left) The smallest inhomogeneous cycle consisting of two 2-degree and two 3-degree sites. (right) Increasing number of interior edges for pentagonal skeleton cycles from zero (simple closed chain) to five (a clique). The presence of interior edges ruins the symmetry of both the weak cycle and the clique such that the nodes are no longer equivalent. 
} \label{fig:diamond}
\end{center}
\end{figure}

We will now demonstrate how to apply this model to determine the percolation properties of inhomogeneous clusters. The smallest inhomogeneous cluster we can study is the 4-node cycle in Fig. \ref{fig:diamond} (left). This cluster has two types of nodes: 2-degree and 3-degree sites. As such, it will require two $g_\tau$ equations to describe its percolation properties. Including tree-like edges, we have the following topology vector
\begin{equation}
    \bm \tau = \{\bot,\cin, \vin\}
\end{equation}
where, imagining that the focal node is at the lowermost vertex in the symbol, $\cin$ indicates the $3$-node site and $, \vin$ represents the $2$-node site in the inhomogeneous 4-cycle. 

The probability that a randomly chosen node is not attached to the GCC is generated by writing Eq. \ref{eq:G0} for our chosen topology vector 
\begin{align}
    G_0(g_\bot,g_{\scalebox{.5}{\cin}},g_{\scalebox{.5}{\vin}})\ =&\sum_{k_\bot=0}^\infty \sum_{k_{\scalebox{.5}{\cin}}=0}^\infty \sum_{k_{\scalebox{.5}{\vin}}=0}^\infty p(k_\bot,k_{\scalebox{.5}{\cin}},k_{\scalebox{.5}{\vin}})\nonumber\\
    &\times g_\bot^{k_\bot}g_{\scalebox{.5}{\cin}}^{k_{\scalebox{.5}{\cin}}}g_{\scalebox{.5}{\vin}}^{k_{\scalebox{.5}{\vin}}}
\end{align}
The probability that a node does not become incorporated into the GCC through a tree-like edge is 
\begin{equation}
    g_\bot = u_\bot + (1-u_\bot)(1-\phi)\label{eq:slbot}
\end{equation}
This is the sum of the probabilities that the neighbouring node is either not attached to the GCC with probability $u_\bot$, or is attached but fails to occupy the edge in the percolation process with probability $(1-u_\bot)(1-\phi)$. 

The probabilities of not becoming part of the GCC through the 4-cycle sites, $g_{\scalebox{.5}{\cin}}$ and $g_{\scalebox{.5}{\vin}}$ are given by 
\begin{align}
    g_{\scalebox{.5}{\cin}} =\ &[u_{\scalebox{.5}{\vin}}+(1-u_{\scalebox{.5}{\vin}})(1-\phi)]^2[u_{\scalebox{.5}{\cin}}+(1-u_{\scalebox{.5}{\cin}})(1-\phi)] \nonumber\\
    &- 2(1-u_{\scalebox{.5}{\vin}})(1-\phi)^2u_{\scalebox{.5}{\vin}}u_{\scalebox{.5}{\cin}}\phi^3\nonumber\\
    &-2(1-u_{\scalebox{.5}{\vin}})(1-\phi)u_{\scalebox{.5}{\cin}}\phi^2(1-u_{\scalebox{.5}{\vin}}\phi^2)(1-(1-u_{\scalebox{.5}{\vin}})\phi)\nonumber\\
    &-2(1-u_{\scalebox{.5}{\cin}})(1-\phi)u_{\scalebox{.5}{\vin}}\phi^2(1-u_{\scalebox{.5}{\vin}}\phi^2)(1-(1-u_{\scalebox{.5}{\vin}})\phi)\label{eq:vin}
\end{align}
and 
\begin{align}
    g_{\scalebox{.5}{\vin}} =\ &[u_{\scalebox{.5}{\cin}}+(1-u_{\scalebox{.5}{\cin}})(1-\phi)]^2 \nonumber\\
    &-2(1-\phi)^2(1-u_{\scalebox{.5}{\cin}})u_{\scalebox{.5}{\vin}}u_{\scalebox{.5}{\cin}}\phi^3\nonumber\\
    &-2(1-u_{\scalebox{.5}{\vin}})(1-\phi)^2u_{\scalebox{.5}{\cin}}^2\phi^3\nonumber\\
    &-2(1-u_{\scalebox{.5}{\vin}})u_{\scalebox{.5}{\cin}}\phi^2(1-u_{\scalebox{.5}{\cin}}\phi^2)^2(1-(1-u_{\scalebox{.5}{\cin}})\phi)\nonumber\\
    &-2(1-\phi)(1-u_{\scalebox{.5}{\cin}})u_{\scalebox{.5}{\cin}}\phi^2(1-u_{\scalebox{.5}{\vin}}\phi^2)(1-(1-u_{\scalebox{.5}{\vin}})\phi)\label{eq:cin}
\end{align}
These equations are understood as follows; firstly, we pick a unique node-site in the cycle as the focal node for the cycle under consideration. The first term in its $g_\tau$ equation is the product of probabilities that each direct-contact edge fails to connect it to the GCC. The leading term of $g_{\scalebox{.5}{\cin}}$ is cubic in $[u_\tau+(1-u_\tau)(1-\phi)]$ while $g_{\scalebox{.5}{\vin}}$ is quadratic. The remaining terms capture the probabilities that nodes use cycle-edges to connect the focal node to the GCC. In other words, we examine the complete set of non-self intersecting walks from all nodes (other than the focal node) that terminate at the focal node. When the nodes are equivalent to one another, then any two paths of the same length have equal probability of attaching the focal node to the GCC. However, when the nodes are heterogeneous, the path probabilities are distinct from one another.

For each successful walk from a cycle-node to the focal node, all other permissible walks back to the focal node (and any node in the success path itself) must not be attached by any path other than the success-path under consideration. To incorporate this, the remaining nodes and edges must fail to connect any success-path node, such that the success-path is allowed to be the connecting path.

Once we have the $g_\tau$ equations we must now find $u_\tau$. Each $u_\tau$ probability satisfies a self-consistent Dyson-like equation found by evaluating Eq. \ref{eq:Jacobian} as
\begin{align}
    u_\bot =\ & G_{1,\bot}(g_\bot,g_{\scalebox{.5}{\cin}},g_{\scalebox{.5}{\vin}})\\
    u_{\scalebox{.5}{\cin}} =\ & G_{1,{\scalebox{.5}{\cin}}}(g_\bot,g_{\scalebox{.5}{\cin}},g_{\scalebox{.5}{\vin}})\\
    u_{\scalebox{.5}{\vin}} =\ & G_{1,{\scalebox{.5}{\vin}}}(g_\bot,g_{\scalebox{.5}{\cin}},g_{\scalebox{.5}{\vin}})
\end{align}
and can be solved using fixed-point iteration.
\begin{figure}[ht!]
\begin{center}
\includegraphics[width=0.475\textwidth]{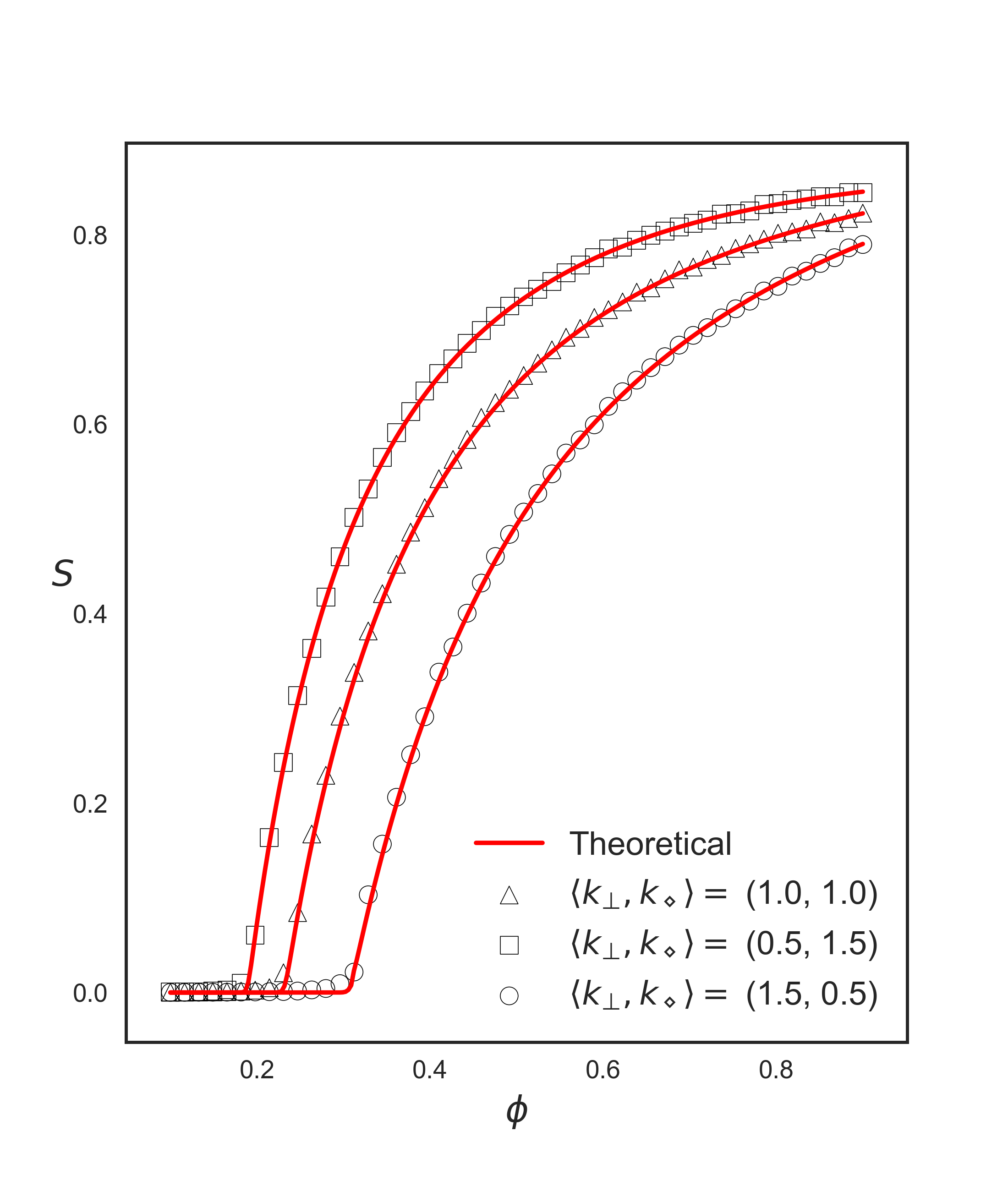}
\caption[diamond]{ 
The percolation properties of random graphs containing the 4-node inhomogeneous cycle and tree-like edges according to Eq \ref{eq:outbreak4} for a variety of clustering vectors with $\langle k_{\scalebox{.5}{\vin}}\rangle=\langle k_{\scalebox{.5}{\cin}}\rangle$. Experimental points are the average of 100 repeats of networks of $N=35000$ nodes.  
} \label{fig:diamond1}
\end{center}
\end{figure}
As a numerical example, consider the case that both the tree-like and the 4-cycle shapes are Poisson distributed, such that the joint degree sequence is 
\begin{equation}
    p(k_\bot,k_\diamond) = e^{-\langle k_\bot\rangle}\frac{\langle k_\bot\rangle^{k_\bot}}{k_\bot!}\left(
    e^{-\langle k_\diamond\rangle}\frac{\langle k_\diamond\rangle^k_\diamond}{k_\diamond!}
    \right)^2
\end{equation}
where $\langle k_{\scalebox{.5}{\cin}}\rangle = \langle k_{\scalebox{.5}{\vin}}\rangle = 2\cdot\langle k_\diamond\rangle / 4$; since, there are 4 nodes in the cycle and 2 nodes in each site. The size of the GCC is found by inserting each $g_\tau$ into Eq \ref{eq:G0} and then subtracting this from one. Following this procedure we find
\begin{equation}
    S = 1 - e^{\langle k_\bot\rangle(g_\bot-1)}e^{\langle k_{\scalebox{.5}{\cin}}\rangle(g_{\scalebox{.5}{\cin}}-1)}e^{\langle k_{\scalebox{.5}{\vin}}\rangle(g_{\scalebox{.5}{\vin}}-1)}\label{eq:outbreak4}
\end{equation}
showing excellent agreement with simulation in Fig.  \ref{fig:diamond1}. 
Consider also the case where the joint degree distribution consists of products of exponentially decaying terms in the number of shapes a node is part of as site $\tau$.
\begin{equation}
    p(k_1,\dots ,k_\nu) = C\cdot e^{-\lambda_1k_1} \cdots e^{-\lambda_\nu k_\nu}
\end{equation}
then the generating function is 
\begin{equation}
    G_0(x_1,\dots,x_\nu) = C\cdot\prod_{\tau\in\bm\tau}  \left(\sum_{k_\tau=0}^{\infty} e^{-\lambda_\tau k_\tau}x_\tau^{k_\tau} \right)
\end{equation}
which evaluates to 
\begin{equation}
    G_0(x_1,\dots,x_\nu) = \prod_{\tau}(1-e^{-\lambda_\tau})\left(\frac{1}{1-x_\tau e^{-\lambda_\tau}}\right)
\end{equation}
using $G_0(\bm 1)=1$ to find $C$. Unlike the Poisson example, we must explicitly construct each $G_{1,\tau}(\bm{x})$ expression according to Eq \ref{eq:Jacobian}. For the exponential degree distribution we find 
\begin{equation}
     G_{1,\nu}(x_1,\dots,x_\nu) =\prod_{\tau\in\bm\tau\backslash\{\nu\}}\frac{(1-e^{-\lambda_\tau})}{(1-x_\tau e^{-\lambda_{\tau}})}\left[\frac{e^{\lambda_\nu}-1}{e^{\lambda_\nu}-x_\nu}\right]^2
\end{equation}

\begin{figure}[ht!]
\begin{center}
\includegraphics[width=0.445\textwidth]{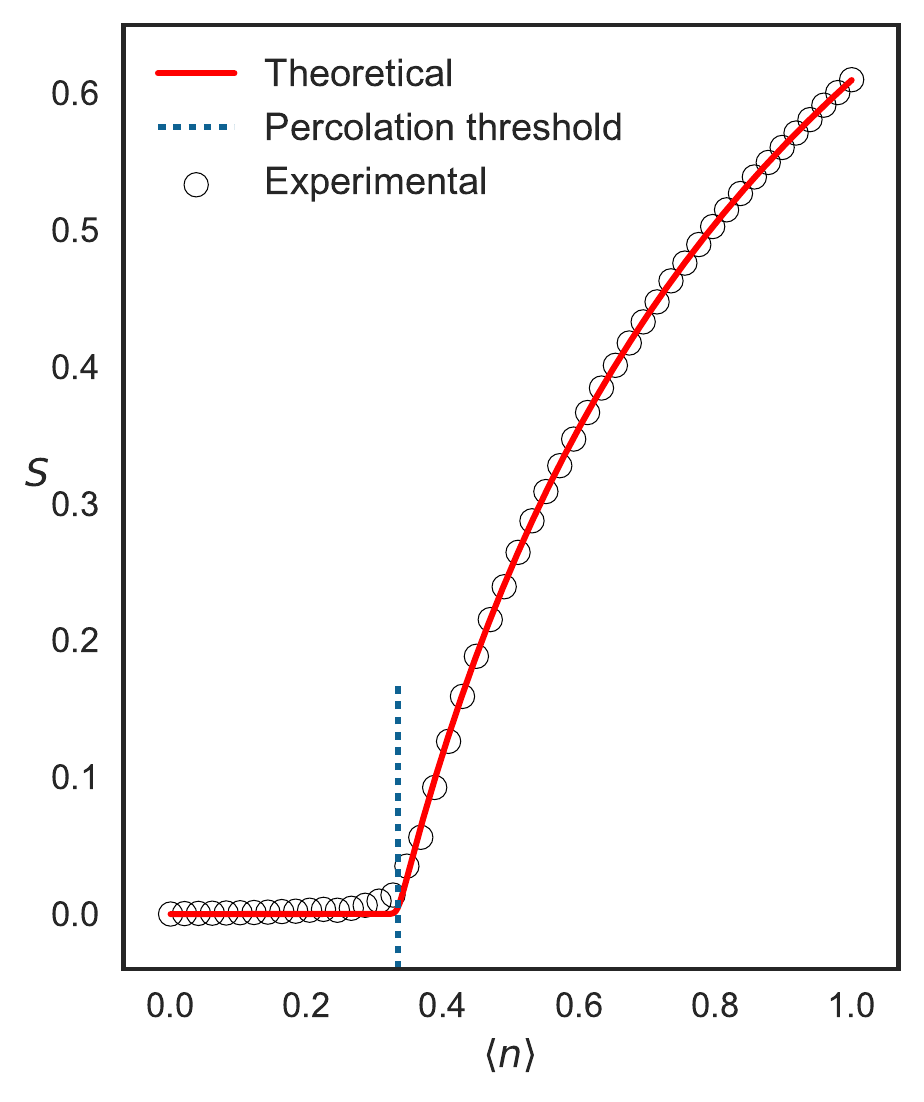}
\caption[diamondmean]{ 
The percolation threshold of a random network of size $N=35000$, $\phi=1$ and with a joint degree distribution given by Eq \ref{eq:pkk}. The critical point is found when $\langle k_{\scalebox{.5}{\vin}}\rangle=1/3$ after which a GCC emerges in the network. The condition arising from the 2-site threshold, $\langle k_{\scalebox{.5}{\cin}} \rangle=1/2$, does not play a role here as a component has already been formed by this point. Intuitively, the GCC forms through the higher-degree site first. Solid lines are theoretical predictions of the model while scatter points are the average result of 100 repeats of percolation over the configuration model networks. 
} \label{fig:diamondphi=1}
\end{center}
\end{figure}

We now explore the particular case when the network contains no tree-like edges, being composed only of the 4-cycle subgraph; and further, we set $\phi=1$. The appropriately generalised Molloy-Reed criterion is found when the following determinant vanishes
\begin{equation}
    \det
    \begin{pmatrix}
    1-\dfrac{2}{\langle n_{\scalebox{.5}{\vin}} \rangle} \dfrac{\partial ^2G_0}{\partial z_{\scalebox{.5}{\vin}}^2} & \dfrac{2}{\langle n_{\scalebox{.5}{\vin}} \rangle} \dfrac{\partial ^2G_0}{\partial z_{\scalebox{.5}{\vin}}\partial z_{\scalebox{.5}{\cin}}}\\[0.8em]\\
    \dfrac{3}{\langle n_{\scalebox{.5}{\cin}} \rangle} \dfrac{\partial ^2G_0}{\partial z_{\scalebox{.5}{\vin}}\partial z_{\scalebox{.5}{\cin}}} & 1-\dfrac{3}{\langle n_{\scalebox{.5}{\cin}} \rangle} \dfrac{\partial ^2G_0}{\partial z_{\scalebox{.5}{\cin}}^2}\\
    \end{pmatrix}
\end{equation}
where $\langle n_\tau\rangle$ is the number of $\tau$-cycles a node is part of on average. The average $\tau$-degree is related to the average number of $\tau$-cycles through $\tau\langle n_\tau\rangle=\langle k_\tau\rangle$. At this point we find
\begin{align}
      &\left[2\frac{\langle  n_{\scalebox{.5}{\vin}}^2\rangle }{\langle  n_{\scalebox{.5}{\vin}}\rangle} -3\right]
      \left[3\frac{\langle  {n^2_{\scalebox{.5}{\cin }}}\rangle}{\langle  {n_{\scalebox{.5}{\cin }}}\rangle} -4 \right] 
     \leq
 6\frac{\langle  {n_{\scalebox{.5}{\vin}} n_{\scalebox{.5}{\cin }}}\rangle^2}{\langle  {n_{\scalebox{.5}{\cin }}}\rangle\langle {n_{\scalebox{.5}{\vin}}}\rangle} 
\end{align}
If the number of sites that a node is part of is distributed according to a Poisson sequence for each site-type, with equal means, $\langle {n_{\scalebox{.5}{\vin}}}\rangle = \langle{n_{\scalebox{.5}{\cin}}}\rangle$, and we impose that each node is only a constituent part of one site-type, then the joint degree distribution is separable in each site-type and is given by
\begin{align}
    p(k_{\scalebox{.5}{\vin}}, k_{\scalebox{.5}{\cin}}) =\ & e^{-\langle k_{\scalebox{.5}{\cin}}\rangle}\frac{\langle k_{\scalebox{.5}{\cin}}\rangle^{k_{\scalebox{.5}{\cin}}}}{k_{\scalebox{.5}{\cin}}!}\delta_{k_{\scalebox{.5}{\vin}},0}\nonumber\\
    &+ 
e^{-\langle k_{\scalebox{.5}{\vin}}\rangle}\frac{\langle k_{\scalebox{.5}{\vin}}\rangle^{k_{\scalebox{.5}{\vin}}}}{k_{\scalebox{.5}{\vin}}!}\delta_{k_{\scalebox{.5}{\vin}},0}\label{eq:pkk}
\end{align}
with $k_{\scalebox{.5}{\vin}}\in 2\mathbb Z$ and $k_{\scalebox{.5}{\vin}}\in \mathbb Z$. In this case the expectation of mixed degrees vanishes and the percolation threshold is at $\langle k_{\scalebox{.5}{\vin}}\rangle = 1/3$, which is less than unity, see Fig \ref{fig:diamondphi=1}. This indicates that the average 4-cycle degree can be less than one and we will still have a GCC in the network. This result supports Karrer \textit{et al} \cite{karrer_newman_2010} who state that the connections a node has due to its presence in a particular cycle may be more significant to the emergence of the GCC than merely the first-order connections, especially if the subgraph is large. 
When the bond occupation probability is less than 1, the derivatives of $g_\tau$ are important and we must modify this condition to 
\begin{align}
&\bigg[\frac{\partial g_{\scalebox{.5}{\vin}}}{\partial u_{\scalebox{.5}{\vin}}}\frac{\langle n^2_{\scalebox{.5}{\vin}} - n_{\scalebox{.5}{\vin}}\rangle}{\langle n_{\scalebox{.5}{\vin}}\rangle} -1\bigg]
\bigg[\frac{\partial g_{\scalebox{.5}{\cin}}}{\partial u_{\scalebox{.5}{\cin}}}\frac{\langle n^2_{\scalebox{.5}{\cin}} - n_{\scalebox{.5}{\cin}}\rangle}{\langle n_{\scalebox{.5}{\cin}}\rangle} -1\bigg]\nonumber\\
&\leq \frac{\partial g_{\scalebox{.5}{\vin}}}{\partial u_{\scalebox{.5}{\vin}}} \frac{\partial g_{\scalebox{.5}{\cin}}}{\partial u_{\scalebox{.5}{\cin}}} \frac{\langle n_{\scalebox{.5}{\vin}} n_{\scalebox{.5}{\cin}}\rangle ^2}{\langle n_{\scalebox{.5}{\vin}} \rangle \langle n_{\scalebox{.5}{\cin}}\rangle}
\end{align}
at the point $u_{\scalebox{.5}{\vin}} = u_{\scalebox{.5}{\cin}}=1$. These derivatives have the effect of reducing the size of the GCC unless $\phi=1$, in which case we recover the original condition. 

\begin{figure}[ht!]
\begin{center}
\includegraphics[width=0.475\textwidth]{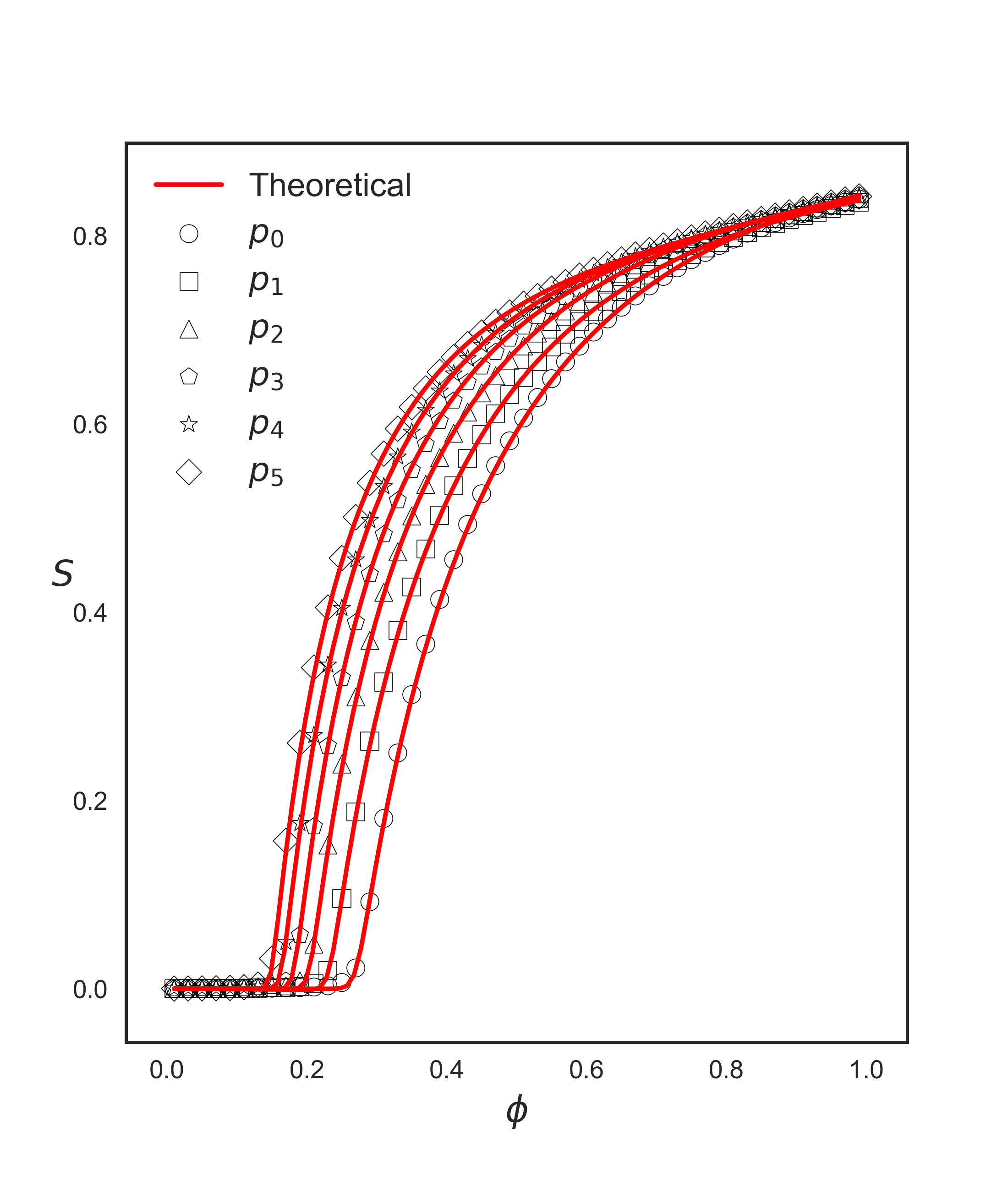}
\caption[pentagonfamily]{ 
The percolation properties of random graphs containing tree-like and pentagonal-skeleton cycles with increasing number of interior-cycle edges according to Fig \ref{fig:diamond} (right). Each topology has a Poisson distributed representation in the joint degree sequence with equal mean degrees set to unity. Theoretical predictions are solid lines while scatter points are the experimental average of 150 repeats of bond percolation over configuration model networks with $N=35000$. The label subscript refers to the number of interior edges in the 5-cycle with $p_0$ being the simple closed chain and $p_5$ being the 5-clique. The $g_\tau$ probabilities and the system of equations are presented in supplementary information to this paper.
} \label{fig:pentagonfamilyperc}
\end{center}
\end{figure}

As a further demonstration of the utility of this method we present the complete system of equations for random graphs consisting of pentagonal backbone cycles in the supplementary material to this paper. Their percolation properties are presented in Fig \ref{fig:pentagonfamilyperc} for an increasing number of interior edges in the 5-cycles. We see excellent matching between theory and experiment in each case. This result shows that the weak-cycle and the clique bound the Molloy-Reed criterion and the GCC fraction for the family of increasing interior-edge cycles. We also observe that at $\phi =1$ the expectation for the GCC converge to the same value. This is because in this limit, the interior edges do not play a role in connecting the cycle to the GCC, this has already been achieved through the cycle outer-skeleton.

With a methodology in place to account for arbitrary inhomogeneous cycles within a network, it remains to  compute each $g_\tau$ equation we require. Realising the increasing complexity of this task as cycles increase in size, we now describe how to write an algorithm to extract the $g_\tau$ equation from a network motif. We support this with a Python code sample in the supporting information. 

In any $g_\tau$ expression we first compute the probability that the direct-contact nodes fail to attach the chosen node to the GCC. We must then subtract from this product the entire set of probabilities that a path other than the direct contact path causes connection to the GCC. This amounts to the enumeration of the walks from each node in a cycle to a defined focal node. The choice of focal node in homogeneous cycles is arbitrary; however, each node must act as the focal node within an inhomogeneous cycle at some point during the calculation, the result of which forms its $g_\tau$ equation as a function of the neighbouring nodes $u_{\bm\tau}$ values.

For a particular cycle and focal node pair, the remaining nodes in the cycle are iterated and a depth-first search can be performed to identify all paths through the cycle from source to focal node target. Each path will contribute to the probability of connecting the focal node to the GCC. To enumerate a specific path, we must count its length and compute its success probability, consisting of
\begin{equation}
u^n\phi^{n-1}(1-u)
\end{equation}
ensuring that the source was attached, each node in the path was unconnected and that bond occupation occurred sufficiently to connect the focal node. The states of the nodes and edges in this path are now determined.

As this path succeeds, all other paths in the cycle must contemporaneously fail to attach the target node. They must also not attach any node in the success path either, as they must be attached only via the success path. Hence, for each remaining node and edge that is in an undetermined state, we must ensure that they fail to change the state of a node or edge in the success path. These probabilities are calculated as one minus the probability that they succeed to attach a specific node in the success path. 

This completes the prescription for calculating the percolation properties of random graphs with arbitrary clustering using this method. 

However, within this automation, it is unclear at what range a node in the graph should be considered as part of a cycle or not considered at all. We suspect that the most accurate model would be to consider the largest Hamiltonian cycle within the network as a single inhomogeneous cluster. Then, compute $g_\tau$ as above for each node within the cycle. However, this is an open question and we welcome any experiments or advice.

\section{2-layer model with closed triples}
\label{sec:2layer}
In addition to symmetry breaking by internal edges, inhomogeneous clusters can also be formed in multilayer networks, where the nodes in a given cycle belong to different layers, with different attributes, despite being degree-equivalent sites. In this section we will consider a bilayer network that extends the formulation of \cite{sbs3887,zhuang_yagan_2016} and exhibits the phase phenomena discovered in \cite{colomer-de-simon_boguna_2014}. This model can be readily generalised to an arbitrary number of layers and cluster types. 

\begin{figure}[ht!]
\begin{center}
\includegraphics[width=0.4\textwidth]{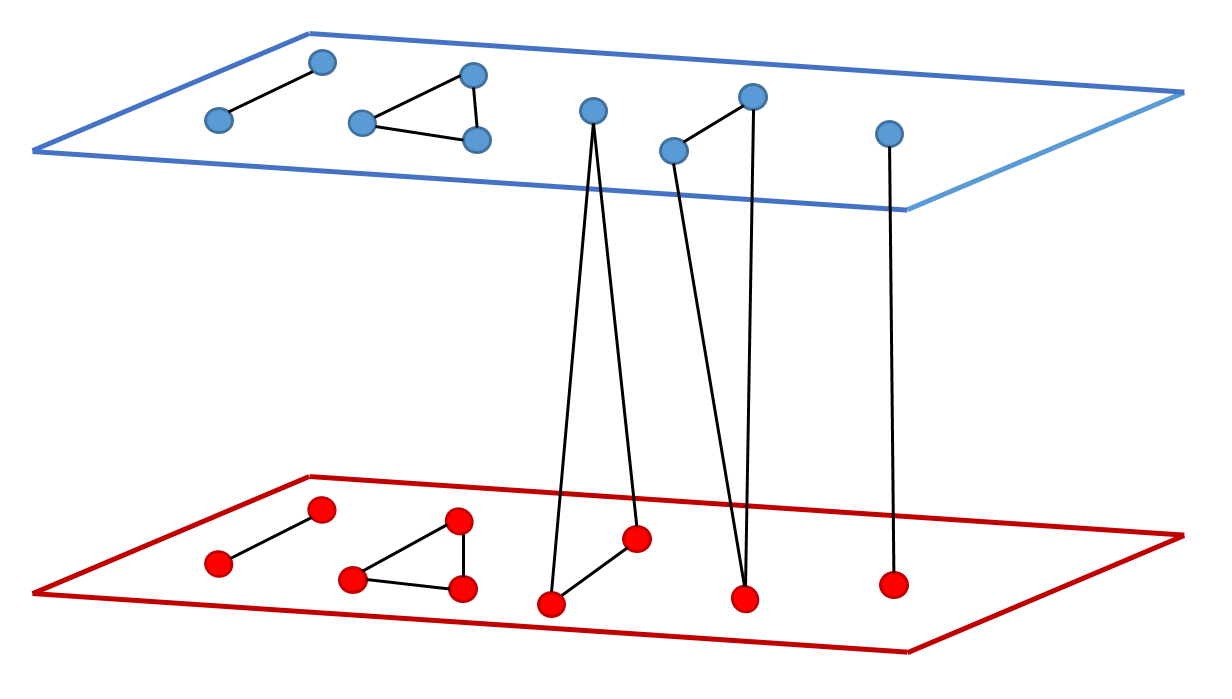}
\caption[topologies]{ 
A visualisation of the vector of topologies in a 2-layer clustered network, $\bm \tau_r = \{ \bot_r,\triangle_r,\blacktriangledown_r, \blacktriangle_r, \top_r  \}$ and correspondingly for the blue layer.
} \label{fig:topologies}
\end{center}
\end{figure}

Consider a 2-layer system consisting of nodes with either red or blue attributes that exhibit clustering in the form of both intra- and inter-layer closed triples along with tree-like edges according to Fig. \ref{fig:topologies}. 
The vector of topologies for each layer can be written as 
\begin{align}
\bm \tau_r =\ & \{ \bot_r,\triangle_r, \blacktriangle_r, \blacktriangledown_r,\top_r  \}\\
\bm \tau_b =\ & \{ \bot_b,\triangle_b,  \blacktriangledown_b,\blacktriangle_b,\top_b  \}
\end{align}
which account for intra-layer tree-like edges, $\bot$, intra-layer triangles, $\triangle$, inter-layer triangles with either a single node in the considered layer, $\blacktriangle$, or two nodes in the considered layer, $\blacktriangledown$, and finally, inter-layer tree-like edges, $\top$. 
The probability of choosing a node at random from the red layer is generated as
\begin{equation}
G_{0,r}(\bm {g_\tau}) = \sum_{k_{\bot,r},\dots,k_{\top,r}=0}^\infty p_r(k_{\bot,r},\dots,k_{\top,r})g_{\bot,r}^{k_{\bot,r}} \cdots g_{\top,r}^{k_{\top,r}} \label{eq:G0_example1}
\end{equation}
and correspondingly for the blue layer
\begin{equation}
G_{0,b}(\bm {g_\tau}) = \sum_{k_{\bot,b},\dots,k_{\top,b}=0}^\infty p_b(k_{\bot,b},\dots,k_{\top,b})g_{\bot,b}^{k_{\bot,b}} \cdots g_{\top,b}^{k_{\top,b}}  \label{eq:G0_example2}
\end{equation}
where $g_\tau$ is the probability that a $\tau$-cycle does not attach the focal node to the GCC. Each $g_\tau$ is a function of the probability that a neighbour node in a $\tau$-cycle is not attached to the GCC, $u_\tau$; we can then write a self-consistent expression for each using Eq. \ref{eq:Jacobian} as 
\begin{equation}
    u_\tau = G_{1,\lambda,\tau} (g_\bot,\dots,g_\top) 
\end{equation}
where
\begin{equation}
    G_{1,\lambda,\tau}(\bm z) = \frac{1}{\langle k_{\lambda,\tau}\rangle}  \frac{\partial}{\partial z_\tau}
    G_{0,\lambda} (\bm z)\label{eq:G12-layerredblue}
\end{equation}
and $\langle k_{\lambda,\tau}\rangle$ is the average number of edges of a node in layer $\lambda$ that are constituent parts of a $\tau$-cycle. For each layer, there are as many excess degree distributions as there are topologies in $\bm {\tau}_\lambda$.

The final step is to compute each $g_\tau$ value for the cycles in the set of cluster topologies which we now turn to examine. For tree-like edges, the probability that the focal node does not become attached to the GCC is the sum of probabilities that the neighbour was itself not attached, $u_\bot$, or that it was attached, $1-u_\bot$, but failed to attach the focal node directly, $1-\phi$, where $\phi$ is the probability of edge occupation. For both intra-layer tree-like edges we have 
\begin{equation}
    g_{r, \bot}(u_{r,\bot}) = [ u_{r,\bot} + (1- u_{r,\bot} )(1-\phi)]\label{eq:rlbot}
\end{equation}
and 
\begin{equation}
    g_{b, \bot} (u_{b,\bot} )= [ u_{b,\bot} + (1- u_{b,\bot} )(1-\phi)]\label{eq:blbot}
\end{equation}
\begin{figure}[ht!]
\begin{center}
\includegraphics[width=0.475\textwidth]{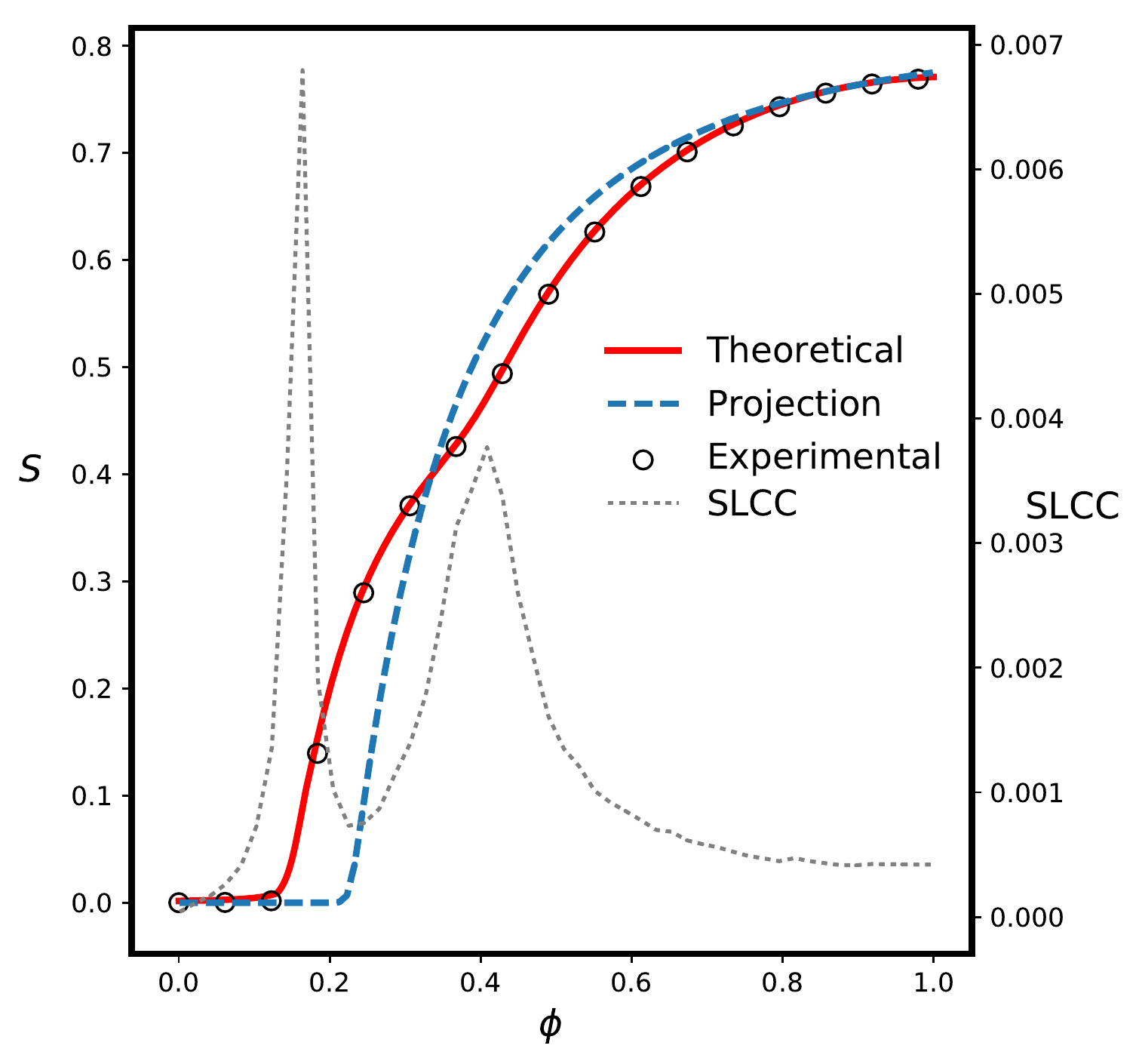}
\caption[numericalexample]{ 
A numerical example of Eqs \ref{eq:2-layerred} and  \ref{eq:2-layerblue} for the clustered 2-layer system exhibiting a double phase transition as well as its monolayer projection formed by aggregating the nodes of both layers. Also plotted is the experimental second largest connected component (SLCC), peaks in which indicate the presence of a phase transition. Bond percolation experiments were conducted on Poisson distributed degree sequences with layers of $N=20000$ nodes each. The mean number of intralayer triangles in the blue layer was set equal to 2, while nodes in the red layer only participate in interlayer triangles. The increased clustering in the blue layer causes the onset of the GCC prior to the phase transition in the red layer, hence we observe the double peak. 
} \label{fig:2_layer}
\end{center}
\end{figure}
Similarly, the intra-layer closed triples are given by 
\begin{align}
    g_{r,\triangle}( u_{r,\triangle}) =\ &  [ u_{r,\triangle} + (1- u_{r,\triangle} )(1-\phi)]^2\nonumber\\
    &- 2\phi^2(1-\phi)(1-u_{r,\triangle})u_{r,\triangle}
\end{align}
and 
\begin{align}
    g_{b,\triangle}( u_{b,\triangle} ) =\ &  [ u_{b,\triangle} + (1- u_{b,\triangle} )(1-\phi)]^2\nonumber\\
    &- 2\phi^2(1-\phi)(1-u_{b,\triangle})u_{b,\triangle}
\end{align}
The inter-layer cycles are composed of more than one type of node and therefore, their $g_\tau$ expressions are functions of more than one argument $u_\nu$ value. The inter-layer tree-like expressions are 
\begin{align}
    g_{r,\top}(u_{b,\top} ) =\ & [ u_{b,\top} + (1- u_{b,\top} )(1-\phi)]\\
    g_{b,\top}(u_{r,\top}) =\ & [ u_{r,\top} + (1- u_{r,\top} )(1-\phi)]
\end{align}
There are four expressions for the inter-layer triangles to compute. First consider the triangle comprising one blue node and two red nodes. This is described by $\blacktriangledown_r$ and $\blacktriangle_b$. Since there is one blue node, the probability that either of the two red nodes do not connect it to the GCC is 
\begin{align}
    g_{b,\blacktriangle}(u_{r,\blacktriangledown}) =\ &  [ u_{r,\blacktriangledown} + (1- u_{r,\blacktriangledown} )(1-\phi)]^2\nonumber\\
    &- 2\phi^2(1-\phi)(1-u_{r,\blacktriangledown})u_{r,\blacktriangledown}
\end{align}
whilst the probability for a red node is given by 
\begin{align}
    g_{r,\blacktriangledown} ( u_{r,\blacktriangledown}, u_{b,\blacktriangle}) = \ & [ u_{r,\blacktriangledown} + (1- u_{r,\blacktriangledown} )(1-\phi)] \nonumber\\
    &- [ u_{b,\blacktriangle} + (1- u_{b,\blacktriangle} )(1-\phi)]\nonumber\\
    &-\phi^2(1-\phi)(1-u_{r,\blacktriangledown})u_{b,\blacktriangle}\nonumber\\
    & -\phi^2(1-\phi)(1-u_{b,\blacktriangle})u_{r,\blacktriangledown}
\end{align}
Similarly, the case where there are two blue nodes and one red node in the inter-layer cluster we have 
\begin{align}
    g_{r,\blacktriangle}(u_{b,\blacktriangledown}) =\ &  [ u_{b,\blacktriangledown} + (1- u_{b,\blacktriangledown} )(1-\phi)]^2\nonumber\\
    &- 2\phi^2(1-\phi)(1-u_{b,\blacktriangledown})u_{b,\blacktriangledown}
\end{align}
whilst the probability for a blue node is given by 
\begin{align}
    g_{b,\blacktriangledown} ( u_{b,\blacktriangledown}, u_{r,\blacktriangle}) = \ & [ u_{b,\blacktriangledown} + (1- u_{b,\blacktriangledown} )(1-\phi)] \nonumber\\
    &+ [ u_{r,\blacktriangle} + (1- u_{r,\blacktriangle} )(1-\phi)]\nonumber\\
    &-\phi^2(1-\phi)(1-u_{b,\blacktriangledown})u_{r,\blacktriangle}\nonumber\\
    & -\phi^2(1-\phi)(1-u_{r,\blacktriangle})u_{b,\blacktriangledown}
\end{align}

If the degree distribution of each topology is Poisson distributed, then the fraction of the GCC on the red layer becomes
\begin{equation}
    S_{r} = 1 - \prod_{\tau \in \bm{\tau_r}} e^{\langle k_\tau\rangle(g_\tau-1)}\label{eq:2-layerred}
\end{equation}
while the blue layer is 
\begin{equation}
    S_{b} = 1 - \prod_{\tau \in \bm{\tau_b}} e^{\langle k_\tau\rangle(g_\tau-1)}\label{eq:2-layerblue}
\end{equation}
Assuming that the layers have the same number of nodes, the total fraction of the GCC for the composite network is then the average of these two quantities, see Fig.  \ref{fig:2_layer}. 
The 2-layer model is created by assigning nodes a colour attribute before endowing them with a degree within each permissible topological edge. The monoplex projection of a multiplex is created by ignoring the colour attribute of the nodes but still retaining the local clustering coefficient. In other words, the model reduces to the tree-triangle model \cite{miller_2009,PhysRevLett.103.058701}, which does not exhibit the double phase transition for uncorrelated joint degree sequences. We will however, revisit this model in section \ref{sec:multiplex} and apply it to the specific case of anticorrelated degree sequences.  

\section{Semi-directed graphs with clustering}\label{sec:directed}

\begin{figure}[ht!]
\begin{center}
\includegraphics[width=0.175\textwidth]{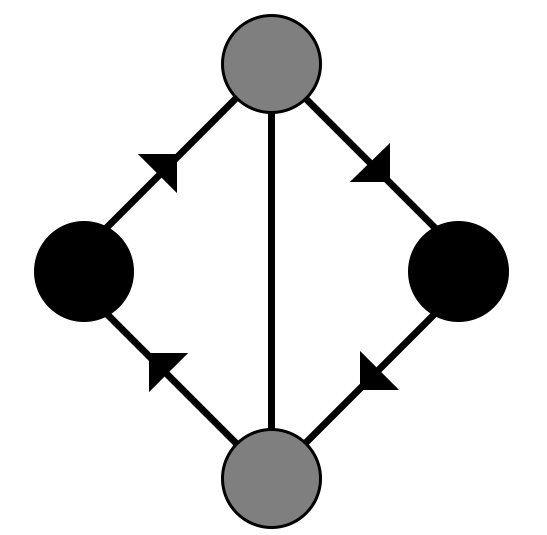}
\caption[directed]{ 
The semi-directed 4-cycle we consider to demonstrate how to approach directed clustered networks. This cycle has 4-fold rotational symmetry meaning that there are still only two unique sites in the cycle. If the interior edge was directed, then the symmetry would be broken and we would have to include an additional site-type in the model.
} \label{fig:directedmotif}
\end{center}
\end{figure}

In this example we will investigate the application of our model to the study of semi-directed networks that exhibit inhomogeneous cycles in their topology using the formulation introduced by Meyers \textit{et al} \cite{meyers_newman_pourbohloul_2006}. As in section \ref{sec:inhomogeneous} we will consider a directed 4-cycle with an undirected interior edge according to Fig \ref{fig:directedmotif}. This cycle has two distinct sites; however, unlike its undirected counterpart, many of the walks back to the focal node are no longer present. Additionally, we split the tree-like degree of a node into tree-like-in degrees, $k_\bot$, and tree-like-out degrees, $k_\top$. Considering these edge-types we have the following generating function
\begin{align}
    G_0(x,y,z,w) =\ & \sum_{ k_\bot=0}^\infty\sum_{k_\top=0}^\infty\sum_{k_{\scalebox{.5}{\vin}}=0}^\infty\sum_{k_{\scalebox{.5}{\cin}}=0}^\infty\nonumber\\
    &\times p(k_\bot,k_\top,k_{\scalebox{.5}{\vin}},k_{\scalebox{.5}{\cin}})x^{k_\bot}y^{k_\top}z^{k_{\scalebox{.5}{\vin}}}w^{k_{\scalebox{.5}{\cin}}}
\end{align}
We then have four excess degree distributions which describe the degree distribution of a node reached by traversing a randomly chosen edge; since there are 4 edges types we may have chosen. Each expression is given by Eq \ref{eq:Jacobian}. We introduce four probabilities, $u_\bot, u_\top, u_{\scalebox{.5}{\vin}}$ and $u_{\scalebox{.5}{\cin}}$ which describe the probability that each site is unattached to the giant strongly connected component (GSCC). 

We can use this system of equations to solve for the size, $S$, of the giant strongly connected component (GSCC) during bond percolation on these networks. To achieve this we have 
\begin{equation}
    S=1-G_0(g_\bot,1,g_{\scalebox{.5}{\vin}},g_{\scalebox{.5}{\cin}})
\end{equation}
where $g_\tau$ is the probability that a $\tau$-edge or cycle fails to connect the focal node to the finite component. Since the network is directed, we notice that none of the $k_\top$ in-edges can become occupied due to their direction, hence we set the variable $y=1$ to prevent counting these edges. 

It remains to derive the expressions for the $g_\tau$ probabilities. The probability that an in-edge fails to attach the node to the percolating cluster is 
\begin{equation}
    g_\bot(u_\bot)=u_\bot+(1-u_\bot)(1-\phi)
\end{equation}
in complete analogy to Eqs \ref{eq:slbot}, \ref{eq:rlbot} and \ref{eq:blbot}. The probability that the 3-site is not attached through the directed 4-cycle is \begin{align}
    g_{\scalebox{.5}{\cin}}(u_{\scalebox{.5}{\cin}},u_{\scalebox{.5}{\vin}})=\ &  
    [u_{\scalebox{.5}{\vin}} + (1-u_{\scalebox{.5}{\vin}})(1-\phi)]\nonumber\\
    &-(1-u_{\scalebox{.5}{\vin}})\phi^2 u_{\scalebox{.5}{\cin}}
    (1-u_{\scalebox{.5}{\vin}}\phi^2)\nonumber\\
    &-(1-u_{\scalebox{.5}{\vin}})u_{\scalebox{.5}{\cin}}u_{\scalebox{.5}{\vin}}(1-\phi)\phi^3\nonumber\\
    &-(1-u_{\scalebox{.5}{\cin}})(1-\phi)u_{\scalebox{.5}{\vin}}\phi^2\nonumber\\
    &-(1-u_{\scalebox{.5}{\cin}})u_{\scalebox{.5}{\vin}}^2\phi(1-\phi)^4
\end{align}
While the probability that the 2-site remains unattached is 
\begin{align}
    g_{\scalebox{.5}{\vin}}(u_{\scalebox{.5}{\cin}},u_{\scalebox{.5}{\vin}})=\ & [u_{\scalebox{.5}{\cin}} +(1-u_{\scalebox{.5}{\cin}})(1-\phi)]\nonumber\\
    &-(1-u_{\scalebox{.5}{\cin}})u_{\scalebox{.5}{\vin}}u_{\scalebox{.5}{\cin}}\phi^3(1-\phi)\nonumber\\
    &-(1-u_{\scalebox{.5}{\cin}})u_{\scalebox{.5}{\cin}}\phi^2 (1-u_{\scalebox{.5}{\vin}}\phi^2)\nonumber\\
    &-(1-u_{\scalebox{.5}{\vin}})u_{\scalebox{.5}{\cin}}\phi^2(1-(1-u_{\scalebox{.5}{\cin}})\phi)
\end{align}
These expressions are quite different to the undirected case in Eqs \ref{eq:vin} and \ref{eq:cin} due to the number of walks that have been lost.
\begin{figure}[ht!]
\begin{center}
\includegraphics[width=0.45\textwidth]{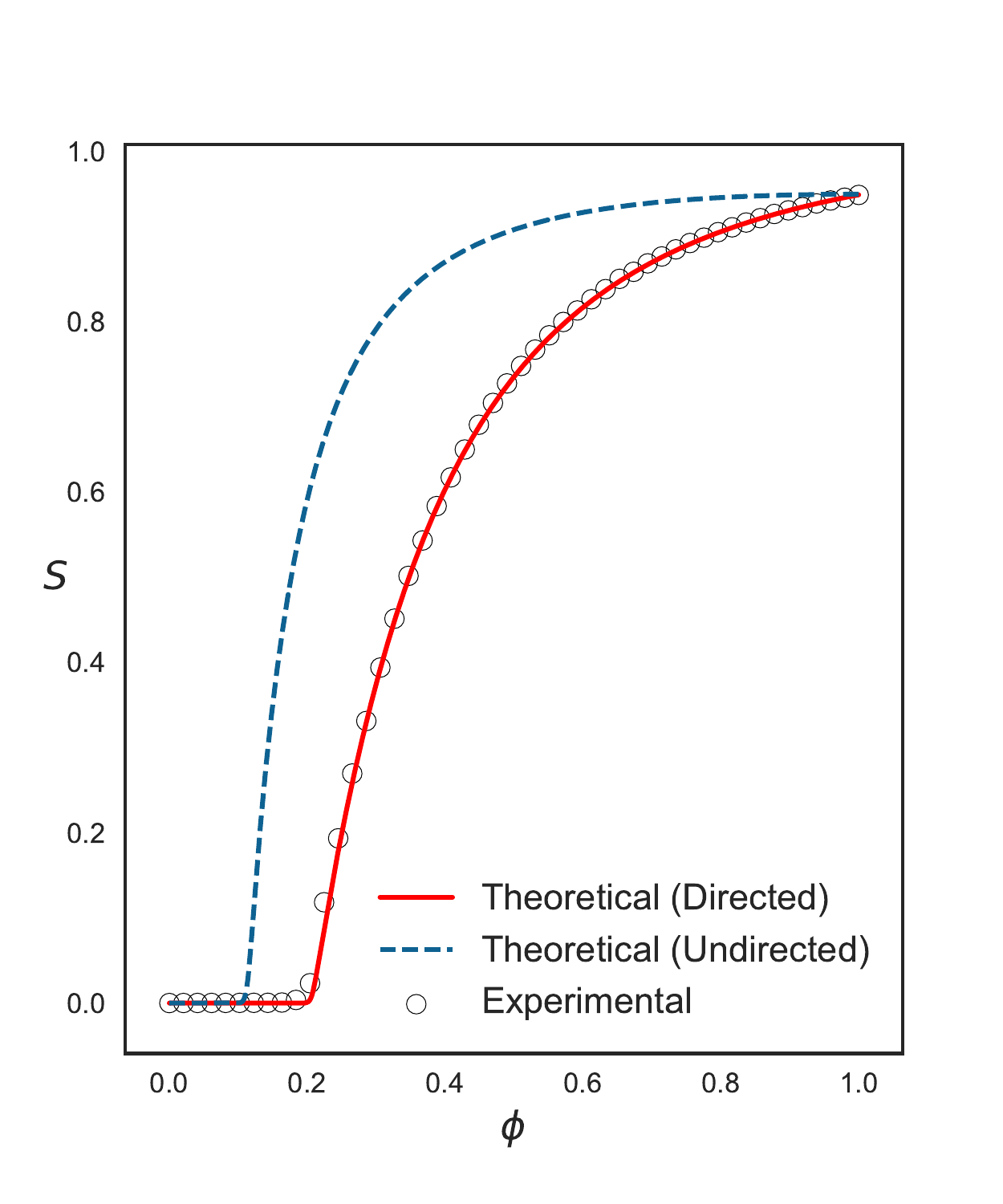}
\caption[directedperc]{ 
The GSCC of random graphs composed of the semi-directed 4-cycle. Circles are experimental, solid red curve is the theoretical model described in section \ref{sec:directed} while the dashed blue line is the theoretical prediction from the undirected inhomogeneous 4-cycle discussed in section \ref{sec:inhomogeneous}. In this experiment, we can see that the undirected network percolates at a lower value of $\phi$ compared to the directed graph. This is because the undirected network has more walks through which attachment to the giant component can occur. The number of tree-like edges was set to zero and the average 4-cycle degree is $\langle k_{\scalebox{.5}{\cin}}\rangle=3$ for both sites.
} \label{fig:directed}
\end{center}
\end{figure}

To complete the calculation we must compute each $u_\tau$ using fixed point iteration to converge on its value in the unit interval. We find 
\begin{equation}
    u_\bot = G_{1,\top}(g_\bot,1,g_{\scalebox{.5}{\cin}},g_{\scalebox{.5}{\vin}})
\end{equation}
which follows occupied edges \textit{backwards} from the focal node to find the connected component. The two 4-cycle quantities are found to be 
\begin{align}
    u_{\scalebox{.5}{\cin}} =& G_{1,{\scalebox{.5}{\cin}}}(g_\bot,1,g_{\scalebox{.5}{\cin}},g_{\scalebox{.5}{\vin}})\nonumber\\
    u_{\scalebox{.5}{\vin}} =& G_{1,{\scalebox{.5}{\vin}}}(g_\bot,1,g_{\scalebox{.5}{\cin}},g_{\scalebox{.5}{\vin}})
\end{align}
Finally, we notice that $u_\top$ does not appear in the system calculation.

\section{Anticorrelated Multiplex networks}
\label{sec:multiplex}

In this final example we will consider multiplex networks with highly anticorrelated degrees \cite{PhysRevX.6.021002}. Multiplex networks are a special class of multilayer network \cite{sbs3887} in which a set of nodes is connected by $M$ different sets of edges. Each layer contains a replicated set of nodes and connects them together with edges of a given type. 

\begin{figure}[ht!]
\begin{center}
\includegraphics[width=0.5\textwidth]{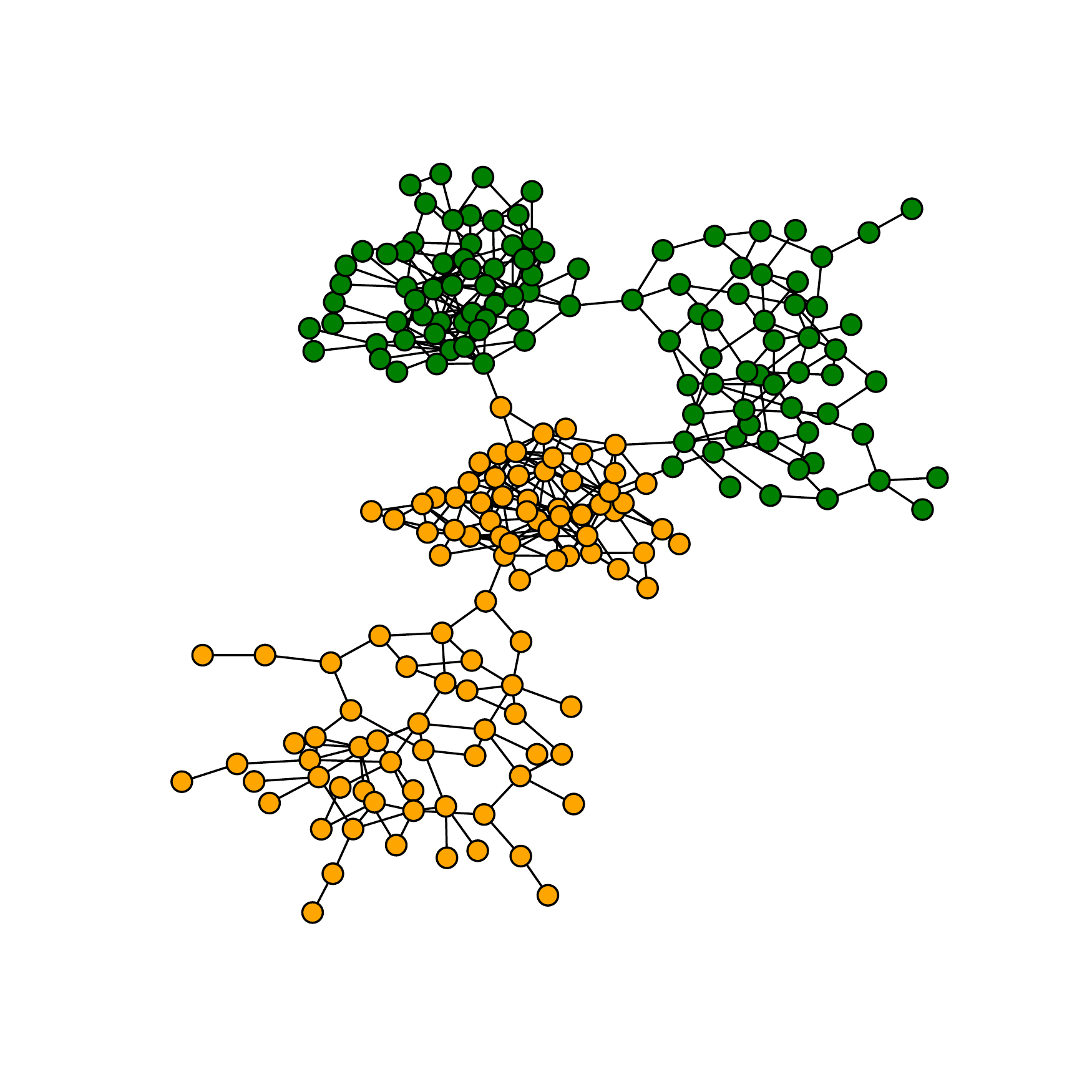}
\caption[orangeGreenmod]{ 
An example of a fully connected multiplex network with two maximally anticorrelated layers composed of tree-like and triangular edge topologies. Each module is sparsely connected together to allow the GCC to span the entire network.  
} \label{fig:orangeGreenmod}
\end{center}
\end{figure}

Let $G$ be a multiplex network consisting of $N$ nodes arranged into 2 edge-layers, one green and the other orange according to Fig \ref{fig:orangeGreenmod}. Each layer contains both tree-like and triangular edges as well as a small number of interlayer edges that allow the GCC to span both layers of the network.

In anticorrelated networks, if a node has an edge of a given colour, then it has a vanishingly small probability of having edges of other colours present. In this model, we extend that property to the anticorrelation of topological edges that each node can be a part of. The joint degree distribution for the maximally anticorrelated degree sequence, with topology vector $\bm \tau = \{1,\dots,n\}$ is given by 
\begin{equation}
p(\bm{k_\tau})=\frac{1}{N}\sum_{\nu\in\tau} N_\nu p(k_\nu) \prod_{\omega\in\bm\tau\backslash\{\nu\}}\delta_{k_\omega,0}
\end{equation}
where $\delta_{i,j}$ is the Kronecker delta. For each $\nu\in \bm\tau$, $k_\nu$ is only non-zero when each $k_\omega$ for $\omega\in \bm\tau\backslash\{\nu\}$ is zero. Note $N_\nu$ is the number nodes in the network that are involved in topology $\nu$. 

When the marginal distribution in each $\nu$ is Poisson distributed with mean degree $\lambda_\nu$, then we have 
\begin{equation}
p(\bm{k_\tau})=\frac{1}{N}\sum_{\nu=1}^n  N_\nu\frac{\lambda_\nu^{k_\nu}e^{-\lambda_\nu}}{k_\nu !}\prod_{\omega\in\bm\tau\backslash\{\nu\}}\delta_{k_\omega,0}
\end{equation}
Then the generating function for the probability of choosing a node at random from the network with a given degree sequence is 
\begin{widetext}
\begin{align}
G_0(z_1,\dots,z_n) =\ &\frac{1}{N}\sum^\infty_{k_1=0}\cdots\sum^\infty_{k_n=0}\left(\sum_{\nu=1}^n N_\nu \frac{\lambda_\nu^{k_\nu}e^{-\lambda_\nu}}{k_\nu !}\prod_{\omega\in\bm\tau\backslash\{\nu\}}\delta_{k_\omega,0}\right)z_1^{k_1}\cdots z_n^{k_n}\\
=\ &\frac{1}{N}\sum^\infty_{k_1,\dots,k_n = 0}\bigg(\sum_{\nu=1}^n N_\nu \frac{\lambda_\nu^{k_\nu}e^{-\lambda_\nu}}{k_\nu !}\prod_{\omega\in\bm\tau\backslash\{\nu\}}\delta_{k_\omega,0}\bigg)\prod_{\mu=1}^n z_\mu^{k_\mu}\\
=\ &\frac{1}{N}\sum_{\nu=1}^n N_\nu\sum^\infty_{k_\nu = 0} \frac{\lambda_\nu^{k_\nu}e^{-\lambda_\nu}}{k_\nu !}z_\nu^{k_\nu}
\end{align}
\end{widetext}


Due to the condition of maximal anticorrelation this expression reduces to 
\begin{equation}
G_0(z_1,\dots,z_n) = \frac{1}{N}\sum_{\nu=1}^n N_\nu e^{\lambda_\nu (z_\nu-1)}\label{eq:G0Anti}
\end{equation}
a term for each topological edge-type in the vector of topologies $\bm \tau$, weighted by the fraction of the graph that it occupies. Under this separable distribution, edges of different colour and topological-type are maximally anticorrelated to one another. The model can be numerically solved by defining a variable, $u_\nu$ for each $\nu\in \bm \tau$, that describes the probability that site $\nu$ remains unattached to the GCC. Each of these variables satisfies a self consistent equation 
\begin{equation}
    u_\nu = \frac{N_\nu}{N}e^{\lambda_\nu(u_\nu^{\kappa_\nu}-1)}
\end{equation}
where $\kappa_\nu$ is the number of edges a site has connecting it the cycle; for instance a node involved in a triangle connects via two edges. Once these variables are found, the percolation properties follow from $S=1-G_0(\bm u_\nu)$. If we relax the strict anticorrelation property, and the intermodule connections are not sparse, then the correct description of the system is given by Eqs \ref{eq:2-layerred} and \ref{eq:2-layerblue} which is the 2-layer model.

\begin{figure}[ht!]
\begin{center}
\includegraphics[width=0.45\textwidth]{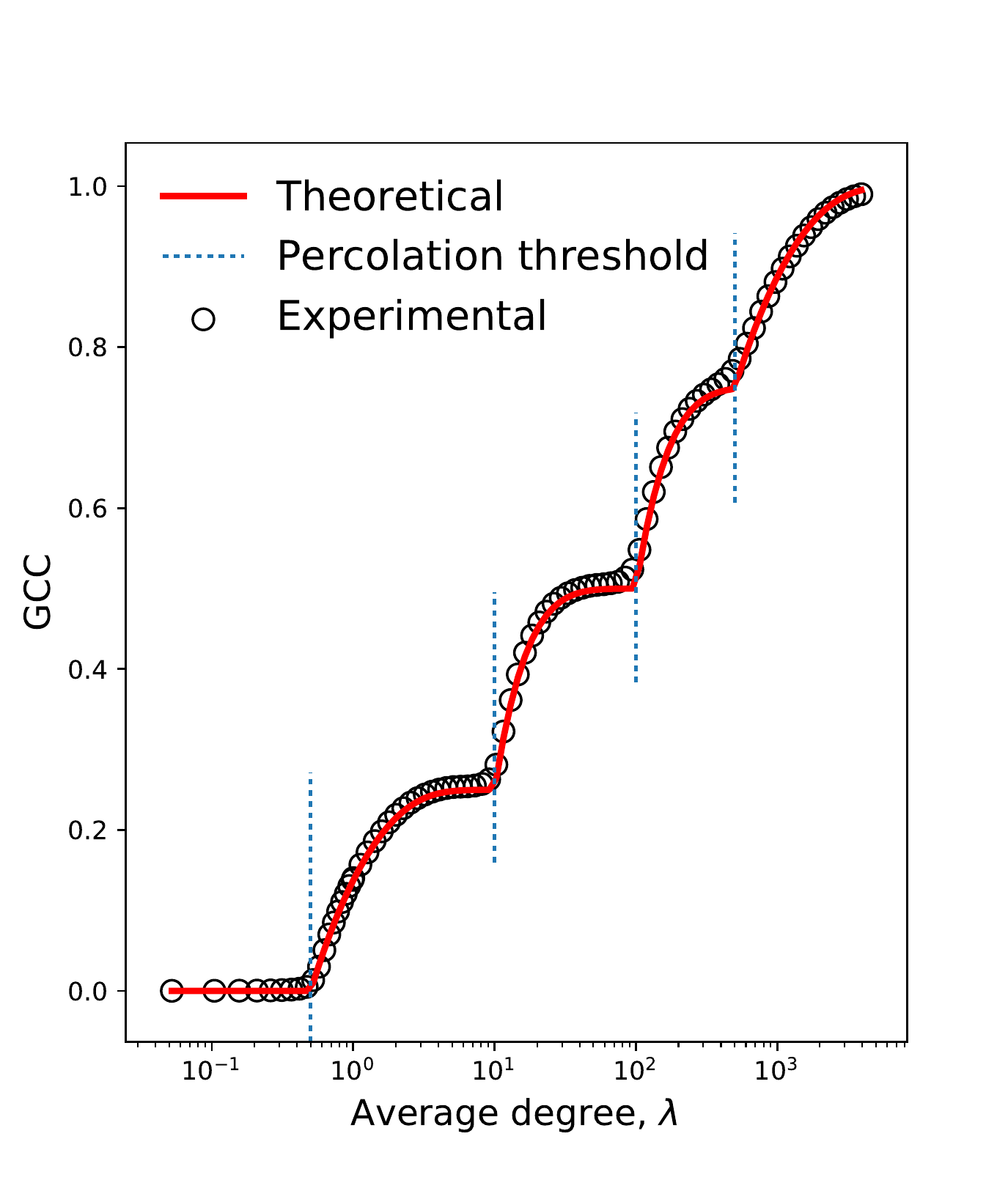}
\caption[orangeGreenexp]{ 
A demonstration of the anticorrelated multilayer clustered network built according to the description in section \ref{sec:multiplex} with $N_\nu=10000$ per layer. The average degree of orange triangles is given by $\lambda$; the average degrees of the remaining topologies have been set such that their percolation transitions occur orders of magnitude apart from one another. Specifically, the Poisson average of the orange tree degrees is $0.1\lambda$, the green triangle average is $0.01\lambda$ whilst the green tree-like edges by $0.001\lambda$. Vertical dashed lines indicate the predicted percolation threshold from Eq \ref{eq:percgen}.
} \label{fig:orangeGreenExp}
\end{center}
\end{figure}

After the construction of the network we have 2-layers each composed of two distinct components: one connected through tree-like edges and the other through triangles in each case. To ensure that the entire network is connected we allow a small number of inter-type edges between both tree-like and triangular components and between the layers of the network itself. The resulting network consists of sparsely connected modules and we observe the percolation properties of each one in Fig \ref{fig:orangeGreenExp}. To highlight the individual contribution each module makes to the overall GCC on the network, we have set the Poisson mean degree of each module to be an order of magnitude apart. This means that the orange triangles, with mean $\lambda$ percolate first, while the orange tree-like module, with mean degree $0.1\lambda$ percolates next. The green tree-like edges follow with mean degree $0.01\lambda$ and finally, the fourth phase transition occurs when the green triangles, with mean degree given by $0.001\lambda$, connect to the GCC. We observe a stepped phase transition for the network as each module connects together. We saw a 2-layer model split into a double phase transition in section \ref{sec:2layer}; now we observe the additional hyperfine splitting of each layer associated with the anticorrelation between degree topologies. 

For each module, the percolation thresholds are given by 
\begin{equation}
    \left((\tau-1)\frac{\langle n_\tau^2\rangle}{\langle n_\tau\rangle}-\tau\right) 
    \leq 0\label{eq:percgen}
\end{equation}
where $\tau$ is the length of the topological cycle and $\langle n_\tau\rangle$ is the average number of cycles a node connects to.

\section{Conclusion}
\label{SEC:conclustion}

In this paper, we have studied the bond percolation process on random networks comprised of arbitrary, inhomogeneous clusters. Through numerical examples we highlighted the application of this formulation to networks composed of inhomogeneous 4-cycles and tree-like degrees. We found excellent matching to experimental simulations of bond percolation as well as the correct conditions for the formation of the GCC. Realising the increasing complexity of elucidating the $g_\tau$ equations for larger cycles, we then described how to algorithmically extract these quantities from a network motif directly. We supported this with supplementary Python code in the supporting information where we describe the family of pentagonal cycles shown in Fig \ref{fig:pentagonfamilyperc}.

Inhomogeneous cycles, under our definition, can also be formed due to nodes belonging to different layers in multilayer networks and so we investigated a 2-layer system with inter- and intralayer clustering. In the case that each layer exhibits different clustering coefficients, we showed a two-point phase transition in the expectation value of the GCC between the two layers, again in agreement with experimental bond percolation. We also supported this with experimental evidence from the SLCC, which showed two peaks, one sharp and the other broad. 

We then investigated a particular example of semi-directed networks, considering again the 4-cycle. We found that the percolation threshold is increased when compared to the undirected case due to the removal of walks through the directed cycle.

In a final example we used an edge-coloured multiplex network to demonstrate a further splitting of the 2-layer clustered model by forcing maximal anticorrelation between degrees of each topology, even within a given edge colour. We found that two clustered layers can exhibit four phase transitions due to each edge topology having different threshold behavior. 

It is hoped that the arguments presented here could help elucidate analytically the percolation properties of empirical networks, an avenue we are keen to investigate.

\subsection*{References}

\bibliography{CLU}

\end{document}